**Current-enhanced broadband THz emission from spintronic devices**

*Mengji Chen, Yang Wu, Yang Liu, Kyusup Lee, Xuepeng Qiu, Pan He, Jiawei Yu, and Hyunsoo Yang\**

M. Chen, Dr. Y. Wu, Y. Liu, Dr. K. Lee, Dr. P. He, Dr. J. Yu, Prof. H. Yang
Department of Electrical and Computer Engineering and NUSNNI-NanoCore
National University of Singapore
117576, Singapore
E-mail: eleyang@nus.edu.sg

Prof. X. Qiu
Shanghai Key Laboratory of Special Artificial Macrostructure Materials and Technology and School of Physics Science and Engineering
Tongji University
Shanghai 200092, China

**An ultra-broadband THz emitter covering a wide range of frequencies from 0.1–10 THz is highly desired for spectroscopy applications. So far, spintronic THz emitters have been proven as one class of efficient THz sources with a broadband spectrum while the performance in the lower THz frequency range (0.1–0.5 THz) limits its applications. In this work, we demonstrate a novel concept of a current-enhanced broad spectrum from spintronic THz emitters combined with semiconductor materials. We observe a 2–3 order enhancement of the THz signals in a lower THz frequency range (0.1–0.5 THz), in addition to a comparable performance at higher frequencies from this hybrid emitter. With a bias current, there is a photoconduction contribution from semiconductor materials, which can be constructively interfered with the THz signals generated from the magnetic heterostructures driven by the inverse spin Hall effect. Our findings push forward the utilization of metallic heterostructures-based THz emitters on the ultra-broadband THz emission spectroscopy.**

Recent technological developments and breakthroughs in terahertz (THz) research have pushed THz technology into a center stage.[1-4] THz technology has been extensively studied for a wide variety of applications, such as THz communications,[5, 6] chemical characterization,[7-9] non-destructive testing,[10, 11] and security surveillance.[12, 13] For these applications, a broadband and efficient THz source is highly desired. In particular, broadening the bandwidth of THz sources while maintaining an excellent dynamic range is important for THz spectroscopies, which require abundant spectrum information.

For THz time domain spectroscopy (TDS), electro-optic (EO) crystals,[14-17] air plasmas,[18, 19] and photoconductive antennas (PCAs),[20-24] are the conventional THz sources for THz pulse generation. However, no single emitter is able to cover a broadband spectrum ranging from 0.1 to 10 THz. At present, PCAs are the most efficient THz emitters for the 0.1–4 THz spectral band while the performance drops dramatically at higher frequencies (> 4 THz).[25-27] Compared with PCAs, some EO crystals and air plasmas have an excellent performance at higher frequencies, whereas the performance in the lower THz frequency range is relatively poor. Thus, combing two different THz emission mechanisms to achieve a broader spectrum has long been thought. However, this approach has been technically limited by the mismatched phase inducing a destructive interference as well as the challenges of heterogeneous integration of different technologies.

During the past several years, the development in the field of spintronics reveals a new possibility for the generation and manipulation of spin and charge currents.[28, 29] Spin-to-charge conversion has been discovered as a new mechanism for ultrafast photocurrent generation. The metallic spintronic THz emitter based on the inverse spin Hall effect (ISHE) is fast becoming a superior candidate among conventional THz sources.[30-35] THz emitters with heavy metal (HM)/ferromagnet (FM) bilayer structures have been demonstrated as broadband THz sources and a gapless spectrum covering up to 10 THz has been reported.[32, 36, 37] On the other hand, the limited performance of HM/FM based THz emitters in the lower

THz frequency range (0.1–0.5 THz) restricts its capacity for a broader range of applications. However, the nanoscale (in terms of thickness) nature of spintronic THz emitter makes it possible to merge with other conventional emitters, while maintaining the similar phase emission leading to a constructive interference of two THz emitters. A negligible dispersion happens when the optical wave propagates through a spintronics device with a thickness in nanometres, and thus no significant phase mismatch occurs between the THz emission from spintronic devices and the other mechanism. Therefore, an ultra-broadband spectra of this hybrid THz emitter can be achieved on a single chip, which has not been explored so far.

In this work, we demonstrate a new type of ultra-broadband spintronic THz emitters assisted by a current modulation on a semiconductor. We observe a significant enhancement of the THz signals in the lower THz frequency range (0.1–0.5 THz), of which the power is 2–3 orders of magnitude larger than that of typical ISHE-based THz emitters. In particular, this enhancement is achieved without compromising the performance at higher frequencies. We further discuss the respective contributions from the magnetic heterostructures and semiconductor materials to THz signals, by alternating the direction of the magnetization with respect to that of bias currents. Our results establish a new method for an extended broadband spintronic THz emitter and thus promote its utilization for a broad range of THz spectroscopies.

A schematic of the expected THz emission from the device is shown in Figure 1a. Under a static external bias, a negligible current flows in the semiconductor without a laser pulse excitation due to the high resistivity of the semiconductor. When a laser pulse with a photon energy larger than the bandgap of the material is incident on the semiconductor channel, electron-hole pairs are created and the resistivity of the semiconductor drops substantially. A transient THz signal is thus emitted due to the ultrafast photoconduction on the semiconductor channel.[20, 38] Meanwhile, the laser pulse simultaneously induces the ultrafast spin current in the FM layer, which then diffuses into the HM layer. Due to the ISHE, the injected spin current is converted into a transverse charge current in the HM layer and induces a THz electromagnetic (EM) radiation from the HM layer.[30, 32-34] It is also worth noting that the thickness of metallic layers is only several nanometers, which is much smaller than the THz wavelength. Therefore, these two near-simultaneously emitted THz signals from the semiconductor and the HM/FM bilayer are in phase and could be constructively combined.

As a proof of concept demonstration, we use a high-resistivity silicon (HR-Si) substrate/Pt or W (6 nm)/Co (3 nm) device, which has a coercivity of ~100 Oe (see Supporting Information Section S1). Figure 1b shows the schematics of the sample structure and the geometric configuration of the device. The metallic thin films are deposited on the 1-mm-thick HR-Si substrate (~ 20 kΩ·cm) by magnetron sputtering with a based pressure less than $5 \times 10^{-9}$ Torr. A typical stack of the magnetic heterostructures consists of a bilayer with 6 nm HM (Pt or W) and 3 nm Co. The thin films are capped with a 4-nm thick $SiO_2$ layer to protect the samples from oxidization. A three-wire pattern (a width of 100 µm, length of 3000 µm and line spacing of 350 µm) with contact pads is fabricated by a process of photolithography and Ar ion beam etching.

In our work, all measurements are performed in a THz time domain spectroscopy (TDS) system (Supporting Information Section S2) at room temperature. An 800 nm *fs* laser pump with a full width half maximum of 120 fs and an energy of 220 µJ is normally incident on the devices from the HM/FM film side and the THz signal is detected after the HR-Si (Figure 1b). A probe beam with an energy of 2 µJ is utilized. A wire grid THz polarizer is applied to define the polarization of the detected THz signals along the *x* axis. In order to minimize current induced heating, we apply a pulsed current with a duration of 20 µs through the wires along the *x* axis in Figure 1b, which is synchronized to the pump laser pulses by a delay generator. An external magnetic field (*H* = 1000 Oe) is applied along the *y* axis, which can saturate the magnetization of the Co layer and maximum the THz emission (Supporting Information Section S1). A current of 100 mA corresponds to a current density of $3.3 \times 10^6$

A/cm$^2$ through the device.

We have carried out the current dependent THz emission measurements and the results are shown in Figure 2. The magnitude of the peak-to-peak THz signal increases, when the positive current density increases. A control experiment is performed, in which there is no laser pulse incident on the devices, and no THz signal is observed for different current densities as shown in the inset of Figure 2a. We also perform the measurements with the HM/FM bilayer grown on a glass substrate, where a semiconductor material HR-Si is absent, and no sizable current dependence on THz emission is observed (Supporting Information Section S3). Therefore, we conclude that THz emission from the HM/FM bilayer is independent of the applied bias current.

A pure ISHE originated THz signal (bias current $I$ = 0 mA) is extracted, as shown in Figure 2b. By subtracting the ISHE originated THz signals in Figure 2b from the current dependent results in Figure 2a, the current dependent signals arising from the semiconductor are extracted in Figure 2c. The THz signal amplitude is proportional to the current density and the THz signal is reversed when the bias current direction is inverted, which is consistent with the photoconduction switching mechanism in semiconductor materials.[39] A current dependent study on a HR-Si/W (6 nm)/Co (3 nm) device is also shown in Figure 2d. The peak-to-peak THz signal increases when a negative current is applied, which is opposite to the results from the Pt based devices in Figure 2a. These two distinct current-dependent results from Pt and W based devices could be attributed to the opposite polarities of the spin Hall angle (SHA) in Pt and W.[40-43] The sign of the THz signals originated from the ISHE is reversed when the SHA of the NM layers has an opposite sign.[34]

We further analyze the data in frequency domain and observe a strong enhancement in a lower frequency range due to the current enhanced mechanism from the HR-Si. The fast Fourier transform (FFT) spectra of the THz signals from the Pt and W based devices are plotted in Figure 3a and 3b, respectively. We show the THz signals under current $I$ = -100, 0 and 100 mA. There is not much different performance of THz signals under three different current densities at higher frequency > 1 THz (see Supporting Information S4 for higher frequency information up to 8 THz). However, in a lower frequency range (0.1–0.5 THz), the magnitude of the THz signals assisted by photoconduction switching with a bias current of 100 mA or -100 mA is up to ~2 orders of magnitude larger than that without a bias current. The insets of Figure 3a and 3b show the contour map of THz signals with respect to the frequency and current density. It is clear that a stronger THz signal is observed in a lower frequency range with increasing the current density. In general, the strongest THz signal from the Pt based device is obtained with positive currents, while the W based device shows the highest output signal when a negative current is applied due to the opposite sign of SHA in W with respect to Pt. The constructive interference of the THz emission from the magnetic heterostructures and the HR-Si significantly extends the bandwidth of our hybrid THz emitters especially in a lower frequency range.

In order to gain a further insight into the respective contribution of the ISHE in the magnetic heterostructures and the photoconduction mechanism in the HR-Si to the THz signals, we measure the current dependent THz signals from a Pt based device with changing the directions of the magnetization and bias current, as shown in Figure 4. Considering the contributions of the ISHE and photoconduction mechanism, the THz output $E(j)$ can be described as $E(j) = E_{\text{semi}}(j) + E_{\text{spin}}$, where $j$ is the current density, $E_{\text{semi}}(j)$ is the current dependent THz signal from HR-Si, and $E_{\text{spin}}$ is the THz signal originating from the spintronic magnetic heterostructures, which is current independent THz signal from the magnetic bilayer. In our experiment, only $E(j)_x$ is measured due to the polarization of the THz polarizer along the $x$ axis, therefore we can write the $x$ component of the THz signal as $E(j)_x = E_{\text{semi}}(j)_x + E_{\text{spin\_}y}$, where $E_{\text{semi}}(j)_x$ represents the THz signal due to the bias current along the $x$ axis and the $E_{\text{spin}\_y}$ is the THz signal due to the external magnetic field along the $y$ axis.

Figure 4a shows the THz TDS signals with an external magnetic field $H$ along the $y$ axis and a current $I$ along the $x$ axis (see Figure 1a for $x$, $y$ and $z$ coordinates), which is the configuration used for Figure 2 and 3. We observe that the THz signals consist of spintronic THz emission, $E_{\text{spin}\_y}$ and semiconductor THz emission, $E_{\text{semi}}(j)_x$. When the $H$ is along the $x$ axis, there is no contribution from the spintronic heterostructures ($E_{\text{spin}\_y} = 0$) and only the current dependent THz signal from the semiconductor ($E_{\text{semi}}(j)_x$) is observed, as shown in Figure 4b. Further, by rotating the device to align the wire pattern parallel to the $y$ axis, we can set the current $I$ along the $y$ axis. When the $H$ is aligned with $I$ along the $y$ axis, the THz signal only originating from the metallic heterostructures is present in Figure 4c ($E_{\text{semi}}(j)_x = 0$). Finally, with the $H$ along the $x$ axis and a bias current along the $y$ axis, no THz signal is observed ($E_{\text{semi}}(j)_x = 0$ and $E_{\text{spin}\_y} = 0$), as shown in Figure 4d. These configuration-dependent results clarify that the THz signal from the magnetic layer is generated only with $H_y$ and that from HR-Si is generated only with $j_x$. Figure 4e-h show the FFT spectra of the THz signals in Figure 4a-4d, respectively. There is an asymmetric enhancement behaviour in a lower frequency region, when sweeping the current (Figure 4e), which was mainly discussed in Figure 2 and 3. However, when the THz signal solely originates from either the semiconductor or the magnetic heterostructures, a symmetric THz response with respect to the current polarity is observed in Figure 4f or 4g, respectively. With $H_x$ and $j_y$ in Figure 4h, no THz signal is observed, in line with the data in Figure 4d.

We have designed a novel ultra-broadband spintronic THz emitter enhanced by a current modulation through the semiconductor channel. The results indicate that the THz emission from this hybrid emitter originates from the combination of the ISHE and the photoconduction mechanism. The ultrafast laser-induced photoconduction in a semiconductor generates a transient photocurrent, filling the gap of the lower THz frequency range. With a current density up to $3.3 \times 10^6$ A/cm$^2$, an enhancement of the THz signals up to 2 orders of magnitude in a lower THz frequency range (0.1–0.5 THz) is achieved. Thus, the bandwidth of spintronic THz emitters can be effectively extended to a lower THz frequency regime. Our results establish a method to optimize and extend the spectrum of spintronic THz emitters and thus push forward the utilization of magnetic heterostructures-based THz emitters on the ultra-broadband THz emission spectroscopy.

## Acknowledgements

M.C. and Y.W. contributed equally to this work. This research is partially supported by the NUS Hybrid-Integrated Flexible (Stretchable) Electronic Systems Program.

[1]     B. Ferguson, X.-C. Zhang, *Nat. Mater.* **2002**, *1*, 26.

[2]     P. H. Siegel, *IEEE Trans. Microwave Theory Tech.* **2002**, *50*, 910.

[3]     M. Tonouchi, *Nat. Photonics* **2007**, *1*, 97.

[4]     S. S. Dhillon, M. S. Vitiello, E. H. Linfield, A. G. Davies, C. H. Matthias, B. John, P. Claudio, M. Gensch, P. Weightman, G. P. Williams, E. Castro-Camus, D. R. S. Cumming, F. Simoens, I. Escorcia-Carranza, J. Grant, L. Stepan, K.-G. Makoto, K. Kuniaki, K. Martin, A. S. Charles, L. C. Tyler, H. Rupert, A. G. Markelz, Z. D. Taylor, P. W. Vincent, J. A. Zeitler, S. Juraj, M. K. Timothy, B. Ellison, S. Rea, P. Goldsmith, B. C. Ken, A. Roger, D. Pardo, P. G. Huggard, V. Krozer, S. Haymen, F. Martyn, R. Cyril, S. Alwyn, S. Andreas, N. Mira, R. Nick, C. Roland, E. C. John, B. J. Michael, *J. Phys. D: Appl. Phys.* **2017**, *50*, 043001.

[5]     J. Federici, L. Moeller, *J. Appl. Phys.* **2010**, *107*, 111101.

[6]     H. J. Song, T. Nagatsuma, *IEEE Trans. Terahertz Sci. Technol.* **2011**, *1*, 256.

[7]     M. v. Exter, C. Fattinger, D. Grischkowsky, *Appl. Phys. Lett.* **1989**, *55*, 337.

[8]     L. Ho, M. Pepper, P. Taday, *Nat. Photonics* **2008**, *2*, 541.

[9]     A. G. Davies, A. D. Burnett, W. Fan, E. H. Linfield, J. E. Cunningham, *Mater. Today* **2008**, *11*, 18.

[10]    P. H. Siegel, *IEEE Trans. Microwave Theory Tech.* **2004**, *52*, 2438.

[11]    M. Yamashita, K. Kawase, C. Otani, T. Kiwa, M. Tonouchi, *Opt. Express* **2005**, *13*, 115.

[12]    K. Kawase, Y. Ogawa, Y. Watanabe, H. Inoue, *Opt. Express* **2003**, *11*, 2549.

[13]    P. U. Jepsen, D. G. Cooke, M. Koch, *Laser Photonics Rev.* **2011**, *5*, 124.

[14]    B. B. Hu, X. C. Zhang, D. H. Auston, P. R. Smith, *Appl. Phys. Lett.* **1990**, *56*, 506.

[15]    Q. Wu, X. C. Zhang, *Appl. Phys. Lett.* **1995**, *67*, 3523.

[16]    J. Hebling, G. Almási, I. Z. Kozma, J. Kuhl, *Opt. Express* **2002**, *10*, 1161.

[17]    X. Mu, I. B. Zotova, Y. J. Ding, *IEEE J. Sel. Top. Quantum Electron.* **2008**, *14*, 315.

[18]   X. Xie, J. Dai, X. C. Zhang, *Phys. Rev. Lett.* **2006**, *96*, 075005.

[19]   K. Y. Kim, J. H. Glownia, A. J. Taylor, G. Rodriguez, *Opt. Express* **2007**, *15*, 4577.

[20]   G. Mourou, C. V. Stancampiano, A. Antonetti, A. Orszag, *Appl. Phys. Lett.* **1981**, *39*, 295.

[21]   G. Mourou, C. V. Stancampiano, D. Blumenthal, *Appl. Phys. Lett.* **1981**, *38*, 470.

[22]   D. H. Auston, K. P. Cheung, P. R. Smith, *Appl. Phys. Lett.* **1984**, *45*, 284.

[23]   E. Sano, T. Shibata, *Appl. Phys. Lett.* **1989**, *55*, 2748.

[24]   M. van Exter, C. Fattinger, D. Grischkowsky, *Opt. Lett.* **1989**, *14*, 1128.

[25]   A. Dreyhaupt, S. Winnerl, T. Dekorsy, M. Helm, *Appl. Phys. Lett.* **2005**, *86*, 121114.

[26]   G. Klatt, R. Gebs, C. Janke, T. Dekorsy, A. Bartels, *Opt. Express* **2009**, *17*, 22847.

[27]   B. Globisch, R. J. B. Dietz, R. B. Kohlhaas, T. Göbel, M. Schell, D. Alcer, M. Semtsiv, W. T. Masselink, *J. Appl. Phys.* **2017**, *121*, 053102.

[28]   A. Hoffmann, *IEEE Transactions on Magnetics* **2013**, *49*, 5172.

[29]   K. Cai, M. Yang, H. Ju, S. Wang, Y. Ji, B. Li, K. W. Edmonds, Y. Sheng, B. Zhang, N. Zhang, S. Liu, H. Zheng, K. Wang, *Nat. Mater.* **2017**, *16*, 712.

[30]   T. Kampfrath, M. Battiato, P. Maldonado, G. Eilers, J. Nötzold, S. Mährlein, V. Zbarsky, F. Freimuth, Y. Mokrousov, S. Blügel, M. Wolf, I. Radu, P. M. Oppeneer, M. Münzenberg, *Nat. Nanotech.* **2013**, *8*, 256.

[31]   T. J. Huisman, R. V. Mikhaylovskiy, J. D. Costa, F. Freimuth, E. Paz, J. Ventura, P. P. Freitas, S. Blügel, Y. Mokrousov, T. Rasing, A. V. Kimel, *Nat. Nanotech.* **2016**, *11*, 455.

[32]   T. Seifert, S. Jaiswal, U. Martens, J. Hannegan, L. Braun, P. Maldonado, F. Freimuth, A. Kronenberg, J. Henrizi, I. Radu, E. Beaurepaire, Y. Mokrousov, P. M. Oppeneer, M. Jourdan, G. Jakob, D. Turchinovich, L. M. Hayden, M. Wolf, M. Münzenberg, M. Kläui, T. Kampfrath, *Nat. Photonics* **2016**, *10*, 483.

[33]  Y. Wu, M. Elyasi, X. Qiu, M. Chen, Y. Liu, L. Ke, H. Yang, *Adv. Mater.* **2017**, *29*, 1603031.

[34]  D. Yang, J. Liang, C. Zhou, L. Sun, R. Zheng, S. Luo, Y. Wu, J. Qi, *Adv. Opt. Mater.* **2016**, *4*, 1944.

[35]  M. Chen, R. Mishra, Y. Wu, K. Lee, H. Yang, *Adv. Opt. Mater.* **2018**, *6*, 1800430.

[36]  T. Seifert, S. Jaiswal, M. Sajadi, G. Jakob, S. Winnerl, M. Wolf, M. Kläui, T. Kampfrath, *Appl. Phys. Lett.* **2017**, *110*, 252402.

[37]  H. S. Qiu, K. Kato, K. Hirota, N. Sarukura, M. Yoshimura, M. Nakajima, *Opt. Express* **2018**, *26*, 15247.

[38]  A. Leitenstorfer, S. Hunsche, J. Shah, M. C. Nuss, W. H. Knox, *Phys. Rev. Lett.* **1999**, *82*, 5140.

[39]  J. T. Darrow, X. C. Zhang, D. H. Auston, J. D. Morse, *IEEE J. Quantum Electron.* **1992**, *28*, 1607.

[40]  L. Liu, O. J. Lee, T. J. Gudmundsen, D. C. Ralph, R. A. Buhrman, *Phys. Rev. Lett.* **2012**, *109*, 096602.

[41]  S. Emori, U. Bauer, S.-M. Ahn, E. Martinez, G. S. D. Beach, *Nat. Mater.* **2013**, *12*, 611.

[42]  H. L. Wang, C. H. Du, Y. Pu, R. Adur, P. C. Hammel, F. Y. Yang, *Phys. Rev. Lett.* **2014**, *112*, 197201.

[43]  Y. Wang, P. Deorani, X. Qiu, J. H. Kwon, H. Yang, *Appl. Phys. Lett.* **2014**, *105*, 152412.

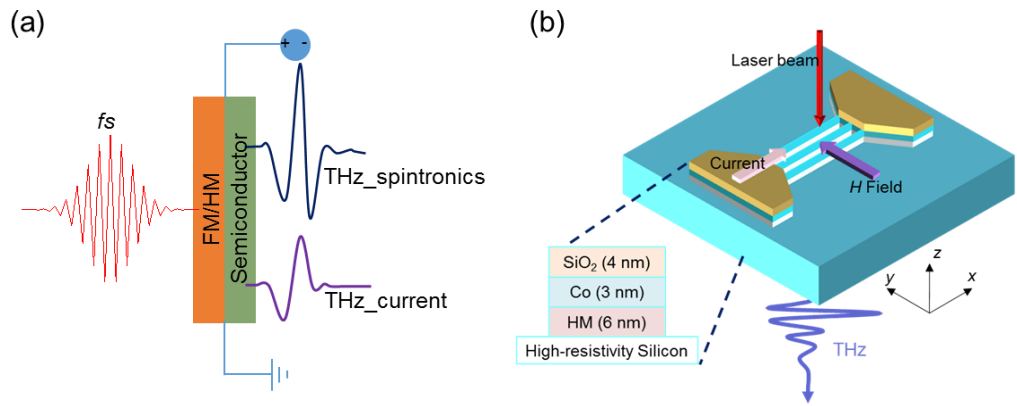

**Figure 1.** (a) A schematic of the THz emission mechanism from the hybrid emitters. An external bias modulation is applied. 'THz_spintronics' refers to THz emission due to the magnetic bilayer, and 'THz_current' refers to the THz output from the semiconductor. (b) Schematics of the sample structure and the geometric configuration of the measurement. An external magnetic field ($H \approx 1000$ Oe) is applied along the $y$ axis and an external bias current is applied along the $x$ axis.

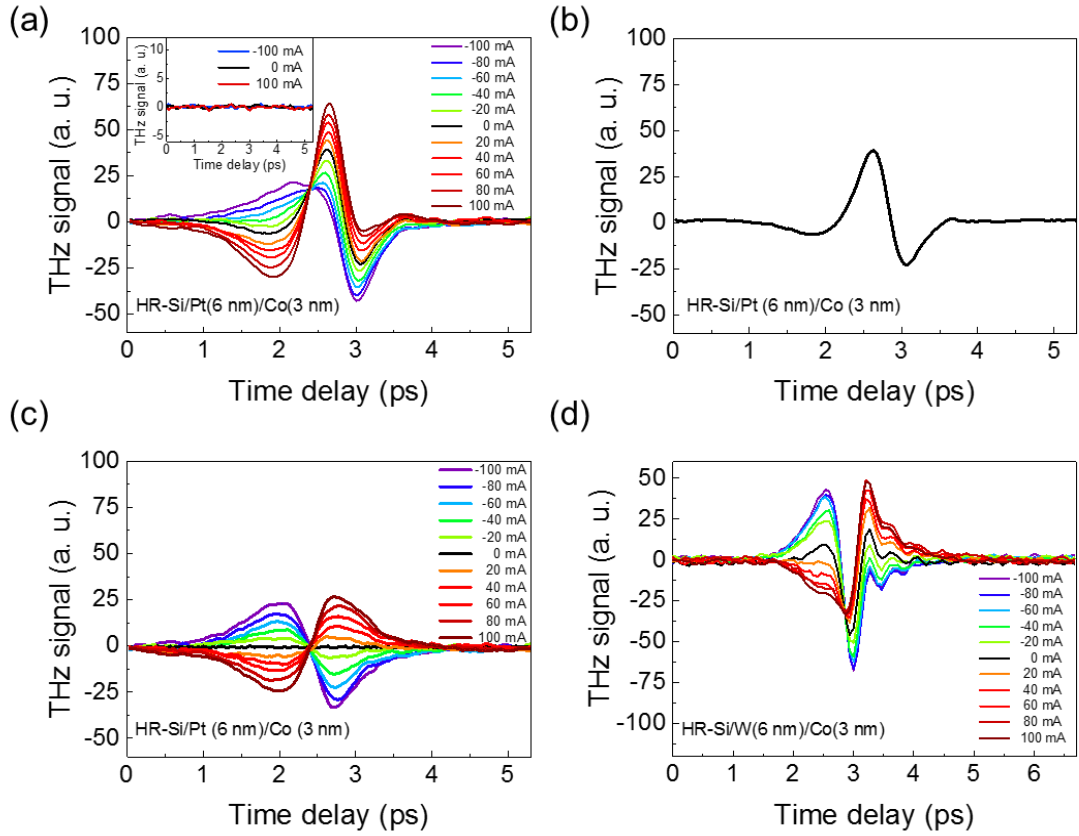

**Figure 2.** (a) Current dependence of THz signals from a HR-Si/Pt (6 nm)/Co (3 nm) device. The inset shows THz signals without a laser pump. With a current of 100 mA, a current density of $3.3 \times 10^6$ A/cm$^2$ is achieved. (b) A typical THz signal from a HR-Si/Pt (6 nm)/Co (3 nm) device without a bias current, which is purely originated from the ISHE. (c) Current dependence of THz signals from the HR-Si only, by subtracting the contribution of the ISHE from (a). (d) Current dependence of THz signals from a HR-Si/W (6 nm)/Co (3 nm) device.

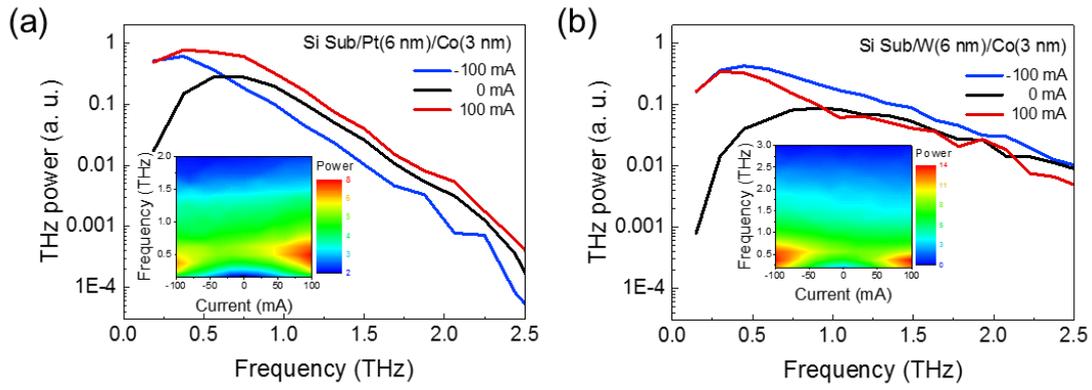

**Figure 3.** Current dependences of the THz FFT signals from (a) a HR-Si/Pt (6 nm)/Co (3 nm) device and (b) a HR-Si/W (6 nm)/Co (3 nm) device. The insets show the THz signal contour maps with respect to the frequency and current. A current of 100 mA corresponds to a current density of $3.3 \times 10^6$ A/cm$^2$.

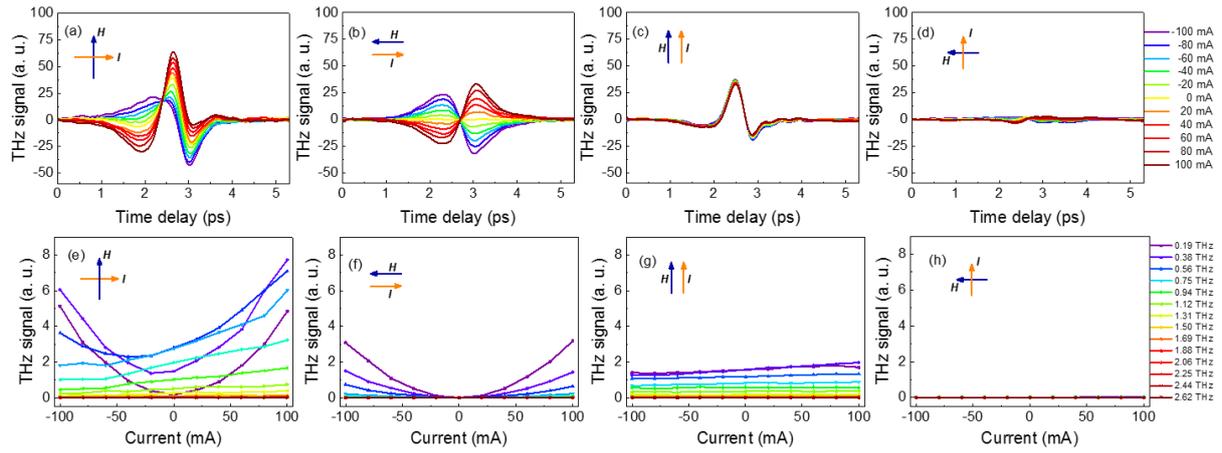

**Figure 4.** Current dependences of (a)-(d) the THz TDS and (e)-(h) the THz FFT spectrum from a HR-Si/Pt (6 nm)/Co (3 nm) device in different measurement configurations: (a) and (e) vertical external magnetic field $H$ (along the $y$ axis) and horizontal bias current $I$ (along the $x$ axis); (b) and (f) horizontal $H$ and $I$; (c) and (g) vertical $H$ and $I$; (d) and (h) horizontal $H$ and vertical $I$. See Figure 1b for $x$, $y$ and $z$ coordinates. A current of 100 mA corresponds to a current density of $3.3 \times 10^6$ A/cm$^2$.

# Supporting Information

**Current-enhanced broadband THz emission from spintronic devices**

*Mengji Chen, Yang Wu, Yang Liu, Kyusup Lee, Xuepeng Qiu, Pan He, Jiawei Yu, and Hyunsoo Yang\**

## S1. Magnetization characterization and magnetic field dependent THz data

Figure S1 shows the magnetic field dependence of the magnetization and THz emission from a high resistivity silicon substrate/W (6 nm)/Co (3 nm). We measure both in-plane and out-of-plane hysteresis loops using vibrating sample magnetometer (VSM), as shown in Figure S1a. A coercivity ($H_c$) of in-plane hysteresis loop is ~ 100 Oe. Thus, an external magnetic field of ~ 1000 Oe is enough to saturate the magnetization to the in-plane direction. Figure S1b shows an in-plane magnetic field dependent THz signal, which shows a similar trend to the magnetization loop data. When the external magnetic field exceeds $H_c$, the intensity of the THz signal remains almost constant.

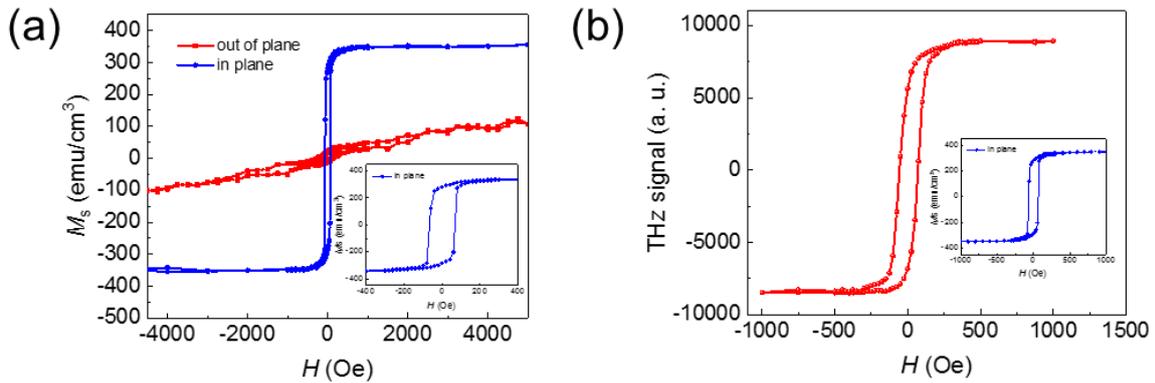

**Figure S1.** (a) In-plane (red) and out-of-plane (blue) *M-H* hysteresis loops of a high-resistivity silicon substrate/W (6 nm)/Co (3 nm) sample. The inset shows an in-plane *M-H* loop from -400 to 400 Oe. (b) THz signal as a function of in-plane magnetic field of a high-resistivity silicon substrate/W (6 nm)/Co (3 nm) sample with the magnetic field from -1000 to 1000 Oe. The inset shows the in-plane *M-H* hysteresis loop from -1000 to 1000 Oe.

## S2. Terahertz time domain spectroscopy system

Figure S2 shows the schematic of the Terahertz time domain spectroscopy (THz TDS) system, which is based on the stroboscopic method. We employ a laser source with a full width at the half maximum of 120 fs, a center wavelength of 800 nm and a repetition rate of 1 kHz. The laser beam is split into two for the stroboscopic sampling. One laser beam with an energy of 220 μJ and a beam diameter of 6 mm is normally incident on the heavy metal/ferromagnetic bilayer, with a linear polarization along the $x$ axis. This pump beam is modulated by a mechanical chopper with a frequecy of 500 Hz. The other laser beam with a much lower energy of 2 μJ and a beam diameter of 2 mm is used for THz signal detection as a probe beam. The emitted THz wave after our device is collected by two parabolic mirror and then focused on a 1-mm thick ZnTe THz detector. The ellipticity of the probe laser beam is modulated by the THz electric field due to the electro-optical effect in the ZnTe crystal, which can be analyzed by a balanced photodetector. Permanent magnets are mounted in a dipole configuration in order to apply the magnetic field (~1000 Oe) along the $y$ axis. A wire grid THz polarizer is utilized after the sample to define the polarizatin of the detected THz wave. The THz generation and detection parts were enclosed in a dry environment with a humidity level of 1.5% to minimized the absorption of THz wave.

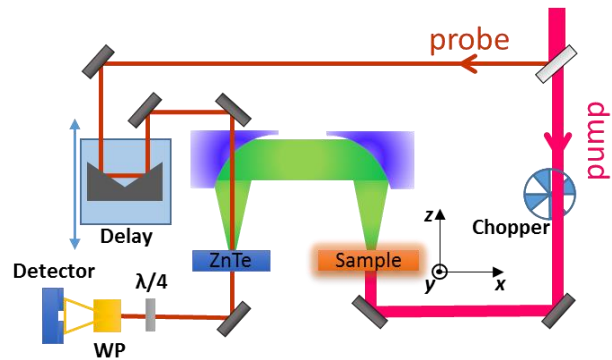

**Figure S2.** The schematic of THz TDS system.

## S3. Current dependent THz signal from glass sub/heavy metal/ferromagnet

Figure S3 shows current dependent measurements of glass substrate/heavy metal (6 nm)/ferromagnet (3 nm). A current density of $3.3 \times 10^6$ A/cm$^2$ is achieved under a bias current of 100 mA. Figure S3a and S3c show the current dependent THz signals in time domain and frequency domain, respectively, from glass substrate/Pt (6 nm)/Co (3 nm). We observe that the THz signals under current $I$ = -100 mA and 100 mA are almost the same as that under $I$ = 0 mA. Especially, in frequency domain, there is no enhancement of THz signal at sub THz range. Similarly, it is observed that the THz signals from glass substrate/W (6 nm)/Co (3 nm) are independent of currents. Therefore, there is negligible current dependent signal in the magnetic heterostructure, and the observed current induced effect can be attributed to the semiconductor layer.

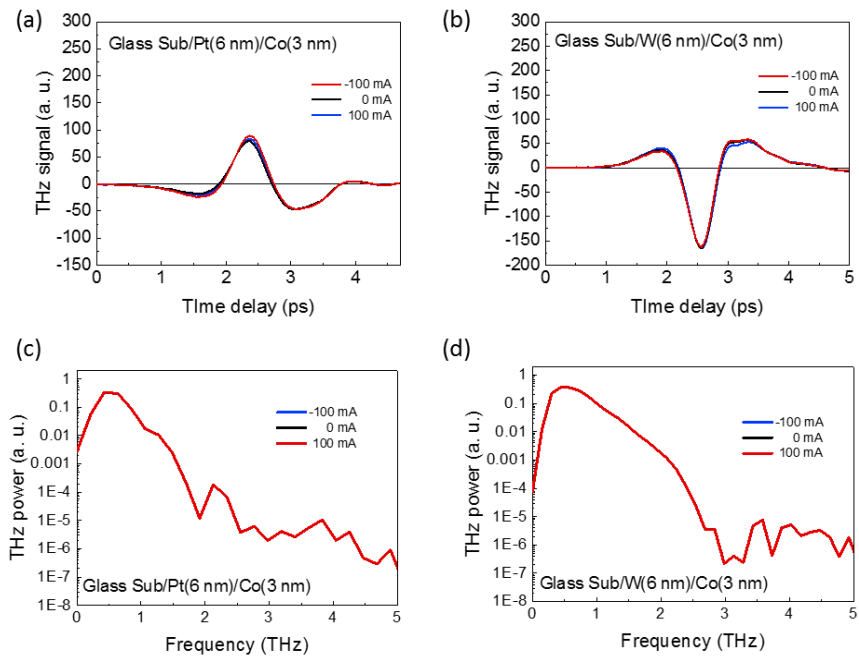

**Figure S3.** Current dependence of THz signal in time domain from (a) glass substrate/Pt (6 nm)/Co (3 nm) and (b) glass substrate/W (6 nm)/Co (3 nm). (c) Current dependence of THz FFT signal of (a). (d) Current dependence of THz FFT signal of (b). A current of 100 mA corresponds to a current density of $3.3 \times 10^6$ A/cm$^2$.

**S4. Frequency domain data for the THz TDS signals**

Figure S4a and Figure S4b show the THz signals in the frequency domain from the Pt and W based device, respectively. Under the current $I$ = -100, 0, and 100 mA, there are comparable performance of THz signals at higher frequency up to 8 THz. A significant enhancement of THz signal happens below 1 THz when the current $I$ = -100 and 100 mA are applied, compared with that without current.

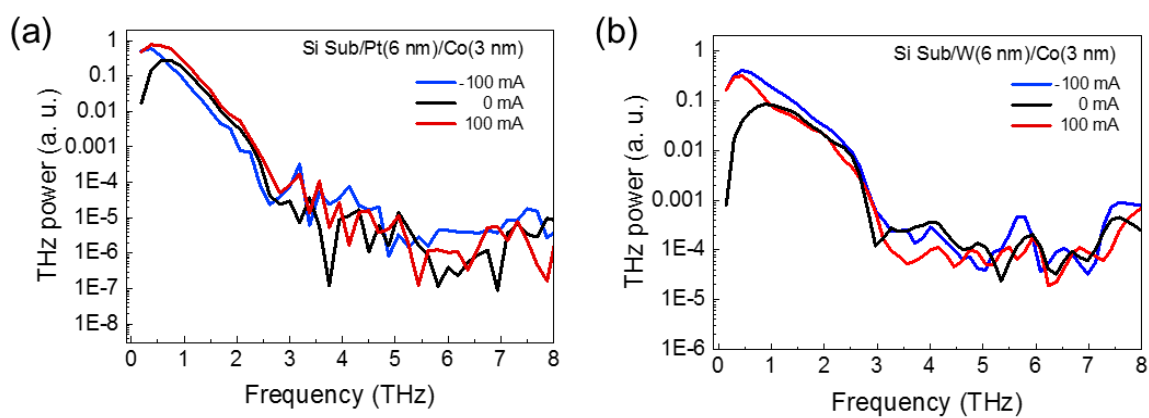

**Figure S4.** Current dependences of the THz FFT signals from (a) a HR-Si/Pt (6 nm)/Co (3 nm) device and (b) a HR-Si/W (6 nm)/Co (3 nm) device. A current of 100 mA corresponds to a current density of $3.3 \times 10^6$ A/cm$^2$.